\documentclass[amsmath,amssymb,lengthcheck,onecolumn,aps,prl] {revtex4}
\usepackage[T1]{fontenc}
\usepackage[utf8]{inputenc}
\usepackage{graphics}
\usepackage{graphicx}
\usepackage{ulem} 
\usepackage{amsmath}

\usepackage{soul}

\usepackage{xcolor}
\begin{document}

\title{ Collective excitation of Bose-Einstein condensate of Bose atoms with P\"oschl-Teller interaction}
\author{Avra Banerjee, Arnab Bhowmik, and Dwipesh Majumder}

\begin{abstract}

We investigate the collective excitations of Bose atoms in the condensate phase with finite-range interactions, modeled using the Pöschl-Teller (PT) potential to explicitly account for the finite interaction range. Utilizing Bogoliubov theory, we derive the excitation spectrum and examine its dependence on interaction parameters. Our analysis reveals that the emergence and disappearance of the roton minima are directly influenced by the range of the PT interaction. Furthermore, we calculate the static structure factor and find enhanced correlations in the low-momentum regime, with their nature controlled by the characteristics of the PT potential. We have examined the interaction dependence of sound velocity and compressibility in detail.
\end{abstract}

\maketitle

\section*{ Introduction}
A gas of Bose particles will undergo a phase transition into a Bose-Einstein condensation (BEC), as was predicted almost 100 years ago \cite{einstein}. The experimental demonstration of Bose-Einstein condensation (BEC) in atomic gases contained in magnetic traps has sparked the interest of researchers \cite{Bose_gas}. Due to the low atomic energies and densities attained in the experiments, an ultracold gas may exhibit behavior in the quantum degenerate domain that dramatically differs from that of an ideal gas, even with weak interactions between its atoms.

The interaction between two Bose atoms at ultracold temperature and low density can be considered as $\delta$--function potential, which can be controlled by Feshbach resonance \cite{fesh,fesh1,fesh2}. The theoretical study of BEC using this potential efficiently explains many aspects of the system. The collective excitation of the BEC in the long wavelength limit is phonon-mode and particle-like excitation in the large wave vector limit, and there is no roton mode in the spectrum with a positive slope at every point. The spectrum is similar, even for the multi-components BEC \cite{roton1,roton4,avra,avra_se,roton_soft}. Some theoretical studies extend the contact interaction to finite-range interaction by considering a hard-sphere potential and found the phonon-roton excitation \cite{roton_soft}. Some interesting studies on spin-orbit coupled BECs may be found here \cite{spinorbit1,spinorbit2}.

Ultracold gases of dipolar particles interacting via long-range anisotropic dipole-dipole potential \cite {DDI1,DDI4} have dramatically changed the nature of the quantum degenerate regimes. 
The dipolar condensate shows the roton-mode of excitations \cite{roton-maxon} which is known in the physics of liquid helium \cite{kapitza,landau,feynman} due to the momentum dependence of the dipole-dipole interaction (DDI). 
The dipolar Bose-Einstein condensates (BECs) have been experimentally achieved using $^{52}$Cr \cite{DDI5}, $^{164}$ Dy \cite{DDI6}, and $^{168}$ Er \cite{DDI7} and consist of atoms with significant magnetic dipole moments. 
The BEC's thermodynamic characteristics, dynamics, and excitations may all be significantly impacted by the DDI \cite{DDI11,DDI12}. Furthermore, they give rise to new quantum phases such as droplet states and supersolid states \cite{supersolid,supersolid1,supersolid2}.

The repulsive symmetric-long range  Coulomb interaction of charged atoms has been the subject of research on the effects of this interaction on the transport properties \cite{coloumb1}, damping \cite{coloumb2}, analytical depletion of the condensates \cite{coloumb3}, and the appearance of the bipolarons \cite{coloumb4}. The collective excitation in this system is monotonic, and there is an absence of a roton-mode of excitation.

The interaction between Bose atoms in the Rydberg-dressed BEC is a symmetric long-range soft-core interaction because of the polarization, in addition to the contact interaction. The excitation of the system has roton-maxon in nature \cite{Ryd_rot, Ryd_rot1}.

The formation of roton minima as a function of interaction potential is fascinating in the Bose gas in the condensed phase.  We have taken into account the  P\"oschl-Teller (PT) \cite{PT1,PT} interaction potential to study the collective excitation in the BEC of uniform density.  The PT potential is an effective model potential for realistic tunable contact interactions in ultracold gases, which has the unique advantage that we can control the range and strength of interaction by changing its parameters \cite{PT_plot}. The interaction parameters can be mapped with the effective scattering length and range of interaction with the actual experiment \cite{aaa}. The PT interaction gives us the opportunity to study the appearance and disappearance of the roton mode by controlling the interaction parameters.


%
The fundamental structure of a BEC could be well approximated by mean-field theory, namely the Gross-Pitaevskii (GP) energy functional. 
Here, we studied the collective excitation using the Bogoliubov approach. It is simple to study the elementary excitation in the uniform BEC using the Bogoliubov theory. 
We have seen how roton modes appear and disappear in the excitation spectra for various interaction strengths and ranges. After that, we studied the structure factors for different temperatures and the interaction of the PT potential.

\section*{ Model and calculation}

In our study, we considered uniform Bose gas at an absolute zero temperature. We have included P\"oschl-Teller interaction (PT) \cite{PT,PT1} potential in addition to the contact interaction between two Bose atoms. The interaction potential of the system can be written as
\begin{equation}
V = U\sum_{i<j} \frac{2\alpha}{cosh^2(\alpha\textbf{r}_{ij})} + g \sum_{i<j} \delta(\textbf{r}_i-\textbf{r}_j),
\end{equation}

 \begin{figure*}
 \centering
  \includegraphics[width=0.9\textwidth]{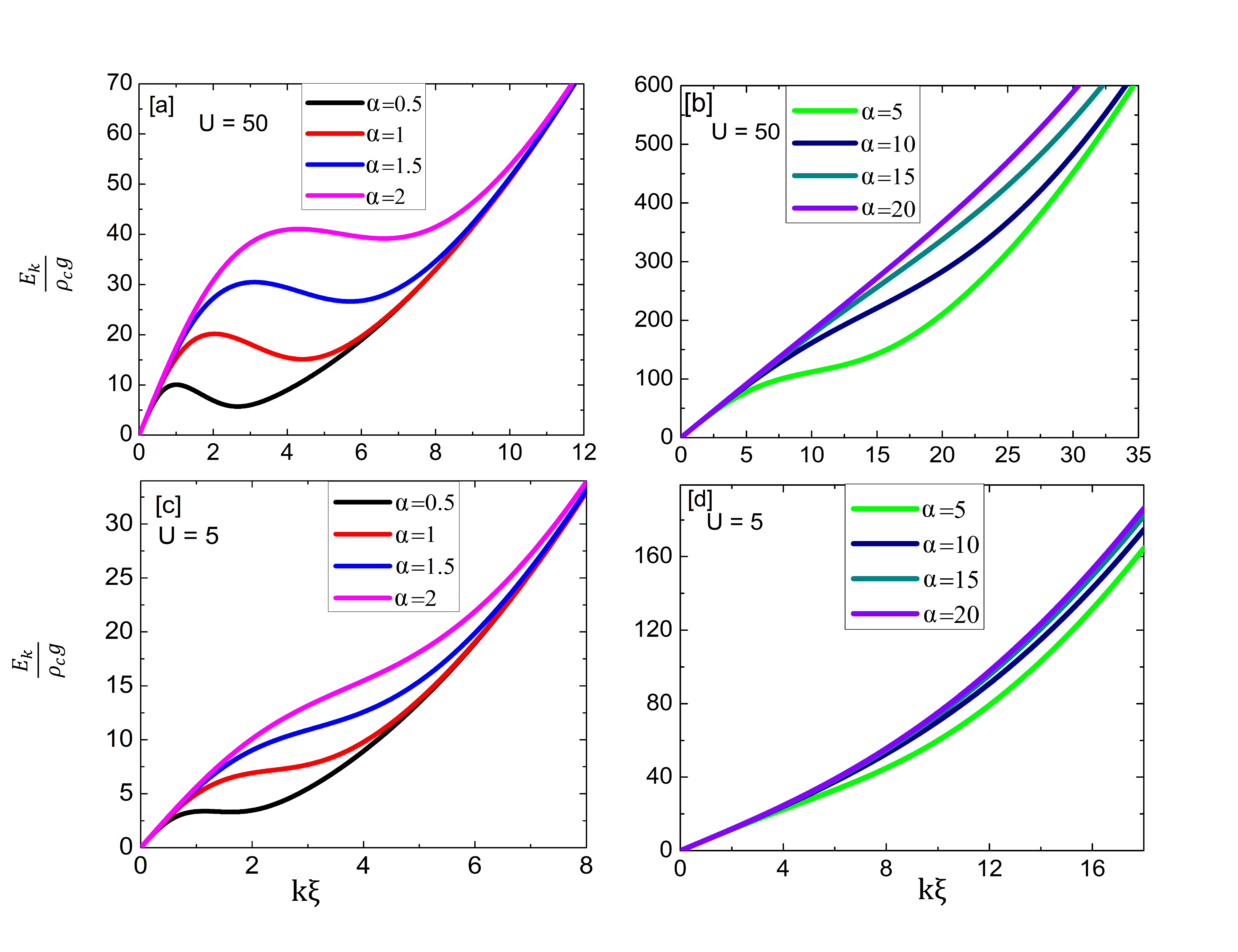}  
  \caption{ Excitation spectra for various values of the parameter $\alpha$ of PT interaction potential. Interaction is a long-range type for smaller values of $\alpha$, so a  roton minima appears. Whereas for larger values of $\alpha$, the interaction becomes a delta function type; that's why the roton minima disappear. When $U$ is smaller, the sharpness of minima is less as contact interaction dominates. \textbf{ Upper panel}: $U=50$, [a] and [b] for different values of $\alpha$, and  \textbf{Lower panel}: $U=5$, [c] and [d] for different values of $\alpha$.   $\alpha$ is expressed in units of $\xi^{-1}$ and $U$ is expressed in units of $\rho_c g \, \xi$.
}
  \label{mu}
   \end{figure*}
   \begin{figure}
  \includegraphics[width=0.5\textwidth]  {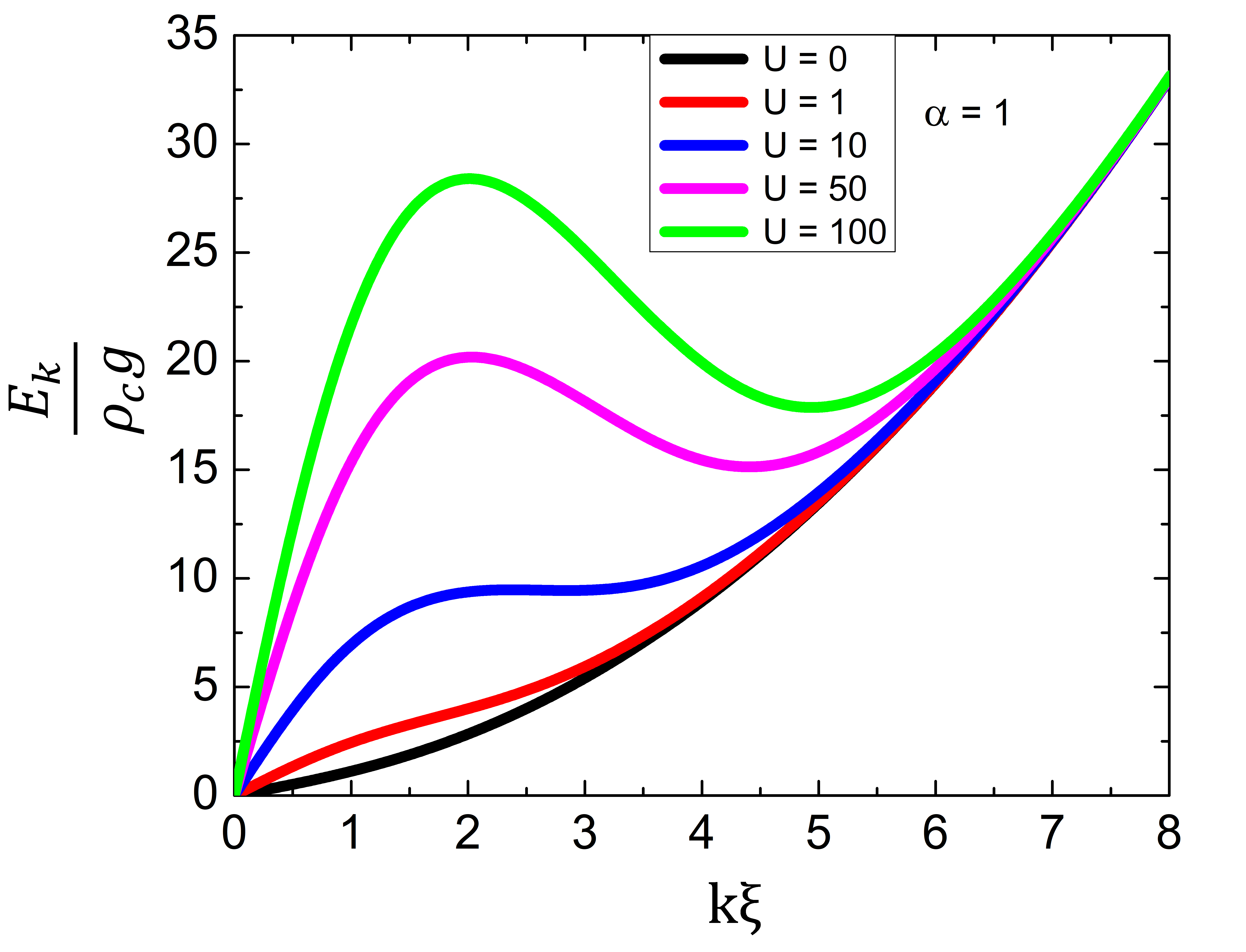}
  \caption{  Excitation spectra for various values of the interaction strength $U$  of the PT interaction potential. For $U=0$, there is only contact interaction between atoms, so there is only a phononic mode of spectra. For larger values of $U$, the strength and range of the PT interaction potential increase. As a result, roton minima start to appear in the spectra. The minima become sharper for larger values of $U$. For lower $U$, the sharpness of the roton minima is less when other parameters are the same.  $\alpha$ is expressed in units of $\xi^{-1}$ and $U$ is expressed in units of $\rho_c g \, \xi$.}
   \label{u}
 \end{figure}
where $U$ is the strength of the P\"oschl-Teller interaction, $\textbf{r}_{ij}$ is the separation distance between the two particles, and $\alpha$ is the interaction parameter expressed as the breadth of the interaction (i.e., $1/\alpha$). A range of $\alpha$ was considered for studying how the excitation spectra of BEC depend on the range of interactions. Here, $U$ is expressed in units of energy $\times$ length ($\rho_c g \, \xi$) and  $\alpha$ is expressed in units of  length inverse ($\xi^{-1}$).
For a very large value of $\alpha$, the potential behaves as a delta function that is contact interaction, whereas if we consider a small value of $\alpha$, the potential becomes long-range (see reference  \cite{PT_plot}). The benefit of this model is that it allows us to switch from a short-range ($\delta$ function) to a long-range interaction, altering the parameter to change the interaction. The second term on the right-hand side is contact interaction. \\


\textbf{Collective Excitations:} We consider a dilute Bose-condensed gas with contact and PT interactions.
The second quantized Hamiltonian of this system \cite{2nd_quan},

\begin{eqnarray}
 H &=& {\displaystyle \int } d^3 \textbf{r} \Big[ {\Psi^\dagger (\textbf{r})\Big(-\frac{\hbar^2\nabla^2}{2m}-\mu\Big)\Psi(\textbf{r})} \nonumber \\
  & &    +\frac{1}{2} {\displaystyle \int } d\textbf{r}'\mathcal{V}(\textbf{r}-\textbf{r}') \Psi^\dagger(\textbf{r}) \Psi^\dagger(\textbf{r}')   \Psi(\textbf{r}') \Psi(\textbf{r}) \Big ],
 \label{Hamiltonian}
\end{eqnarray}


where the first term is the kinetic energy, $\mu$ is the chemical potential, $\Psi(\textbf{r})$ is the field operator, $m$ is the mass of a particle, and the interaction potential $\mathcal{V}(\textbf{r})$ includes the contact and the PT interaction potential, which is given by 

\begin{eqnarray}
  \mathcal{V}(\textbf{r} - \textbf{r}')&=& g\delta(\textbf{r} - \textbf{r}')+ U \frac{2\alpha}{\cosh^2(\alpha\, |\textbf{r} - \textbf{r}'|)}.  
\end{eqnarray}

The system is at zero temperature; almost all the atoms are in the condensate state, and some atoms are in the excited states due to interaction between them, which is effectively fluctuation over the condensate. We split the field operator into two parts, as follows.

\begin{equation}
  \Psi(\textbf{r})=\Phi(\textbf{r})+\varphi (\textbf{r}),
  \label{split_wave}
  \end{equation}
where $\langle\Phi({\bf r}) \rangle$ is the condensate wave function, which is related to the density of the condensate by $\rho_c({\bf r}) = |\Phi({\bf r})|^2$ and  $\varphi({\bf r})$ is the fluctuation $\bigr ($$\langle\varphi({\bf r})\rangle=0$ $\bigr)$, which has been neglected in the mean-field theory.

As the lowest energy state is macroscopically occupied, we can neglect the higher order terms of $\varphi$ in the Hamiltonian (\ref {Hamiltonian}). We put the field  (\ref{split_wave}) in the Hamiltonian, and by considering only quadratic and quartic terms in the noncondensate operators, we get \cite{hamil},

\begin{eqnarray}
  H&=& H_0 +    {\displaystyle \int }  d^3 \textbf{r} \varphi^\dagger (\textbf{r}) \left [-\frac{\hbar^2\nabla^2}{2m}-\mu \right ]\varphi(\textbf{r}) \nonumber \\
            &+&\frac{1}{2}  {\displaystyle \int } d^3\textbf{r} d^3\textbf{r}' \mathcal{V}(\textbf{r}-\textbf{r}') \biggr [\Phi(\textbf{r})\Phi(\textbf{r}')\varphi^\dagger(\textbf{r})\varphi^\dagger(\textbf{r}') \nonumber \\   
           & &+\Phi^*(\textbf{r})\Phi^*(\textbf{r}')\varphi(\textbf{r}')\varphi(\textbf{r})+  \Phi^*(\textbf{r})\Phi(\textbf{r}')\varphi^\dagger(\textbf{r}')\varphi(\textbf{r})  \nonumber \\
  &&  +  \Phi(\textbf{r})\Phi^*(\textbf{r}')\varphi^\dagger(\textbf{r})\varphi(\textbf{r}')+
     \Phi(\textbf{r}')\Phi^*(\textbf{r}')\varphi^\dagger(\textbf{r})\varphi(\textbf{r}) \nonumber \\
  &&  +  \Phi(\textbf{r})\Phi^*(\textbf{r}) \varphi^\dagger(\textbf{r}')\varphi(\textbf{r}')\biggr ], 
   \label{H_nc}
\end{eqnarray}
where $H_0$ is the Hamiltonian of the condensate part of the system

\begin{eqnarray}
  H_0 &=&  {\displaystyle \int }  d^3 \textbf{r} \Phi^* (\textbf{r}) \left [-\frac{\hbar^2\nabla^2}{2m}-\mu \right ]\Phi(\textbf{r}) \nonumber \\
  &&+  \frac{1}{2}  {\displaystyle \int } d^3\textbf{r} d^3\textbf{r}' \mathcal{V}(\textbf{r}-\textbf{r}') \Phi(\textbf{r})\Phi(\textbf{r}')\Phi^\dagger(\textbf{r})\Phi(\textbf{r}') \nonumber .
\end{eqnarray}

 The terms corresponding to the anomalous noncondensate density can be ignored and to its complex conjugate by the Popov approximation \cite{hamil1,hamil2}.
In this instance, we regard $\Phi (\textbf{r})$ to be real without sacrificing generality \cite{hamil3}. Non-condensate part of the Hamiltonian (\ref{H_nc}) may therefore be expressed as follows \cite{hamil,2nd_quan}:

\begin{eqnarray}
  H_{nc} =  &  \biggr . {\displaystyle \int } d^3\textbf{r} \varphi^\dagger(\textbf{r})\mathcal{L}_o\varphi(\textbf{r})+  
  \frac{1}{2}{\displaystyle \int } d^3\textbf{r}   d^3\textbf{r}'\mathcal{V}(\textbf{r}-\textbf{r}')  \nonumber \\
  & 
    \biggr [\rho_c(\textbf{r},\textbf{r}')\Big(\varphi^\dagger(\textbf{r}')\varphi(\textbf{r})+\varphi^\dagger(\textbf{r})\varphi(\textbf{r}') \Big) \nonumber \\
  &  +   \Phi(\textbf{r})\Phi(\textbf{r}')\Big(\varphi^\dagger(\textbf{r})\varphi^\dagger(\textbf{r}')+\varphi(\textbf{r}')\varphi(\textbf{r})\Big)\biggr ] .
    \label{H_nc2}
\end{eqnarray}

Here, we assume that the depletion of the atoms in the excited state is very low, as the density of the gas is low and the temperature is also zero. So, we have the approximation $\rho_t \text{(total density)} \simeq \rho_c \text{(condensate density)} $, and $\mathcal{L}_o$ is given by, 
\begin{eqnarray}
\mathcal{L}_o= -\frac{\hbar^2\nabla^2}{2m}-\mu + {\displaystyle \int }  d^3\textbf{r}'\mathcal{V}(\textbf{r}-\textbf{r}') \rho_c(\textbf{r}') \nonumber .
\end{eqnarray}

 \begin{figure}
  \includegraphics[width=8.8cm,height=7cm]{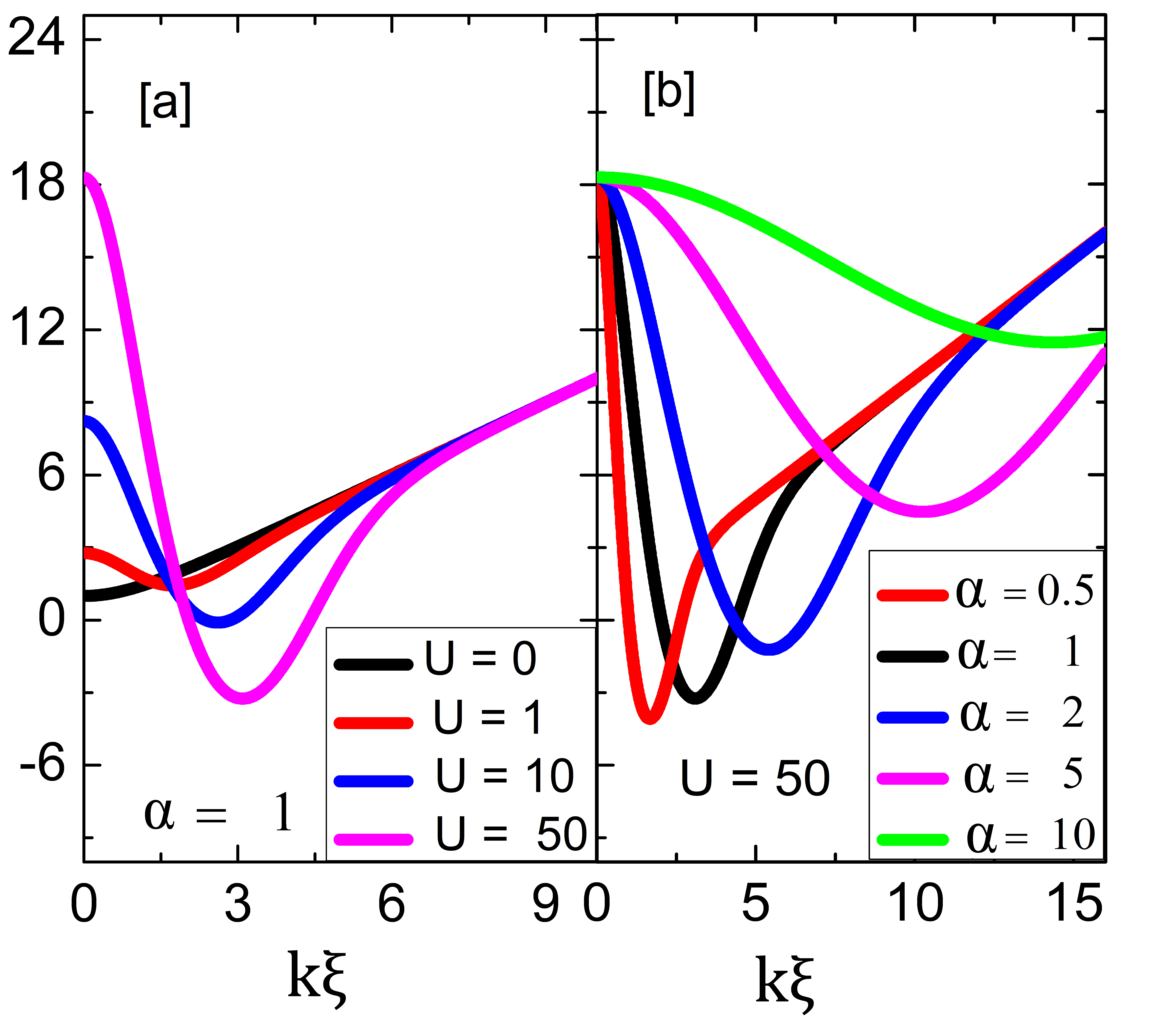}
  \caption{ [a] $\frac{1}{\rho_c g}\frac{dE_k}{d( k\xi)}$ (slope of the excitation spectrum) for different values of $U$ and $\alpha=1$. It is studied that the gap at $k\to0$ increases with increasing $U$, which means the sound velocity increases with $U$. [b]  $\frac{1}{\rho_c g}\frac{dE_k}{d( k\xi)}$ for different values of $\alpha$ and $U=50$. Here, the gap remains unchanged for different values of $\alpha$, which means the sound velocity $(v_s)$ is independent of $\alpha$. A negative slope suggests a roton mode in the excitation spectrum.  }
   \label{du}
 \end{figure}

\begin{figure*}
\includegraphics[width=1.0 \textwidth]{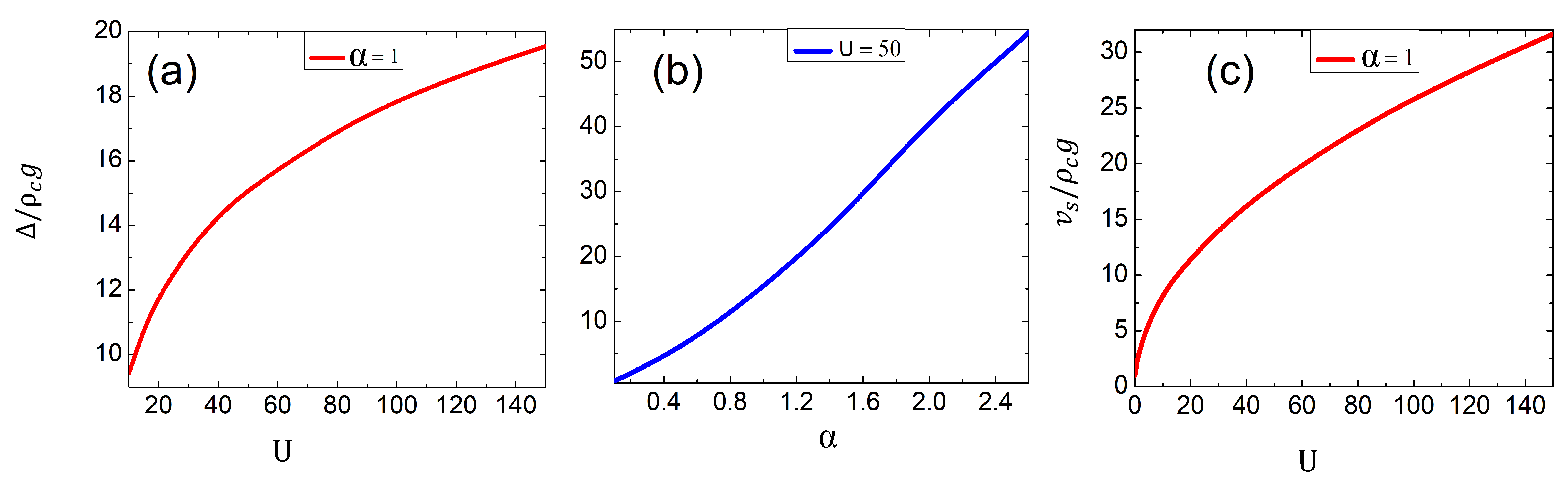}
  \caption{variation of roton minima energy ($\Delta$ denotes the roton minima energy, defined as the lowest value of the excitation energy at the roton momentum region) for different values of $U$ and $\alpha$ is shown in figures (a) and (b), respectively. The roton mode appears for $U > 10$ when $\alpha= 1$, and the roton minima energy increases with $U$. The roton minima energy also increases with $\alpha$, and the roton mode disappears for $\alpha> 2.6$ when $U = 50$. Figure (c) shows that sound velocity $(v_s)$ increases with $U$. The sound velocity does not depend upon $\alpha$.   $\alpha$ is expressed in units of $\xi^{-1}$ and $U$ is expressed in units of $\rho_c g \, \xi$.}
  \label{roton}
 \end{figure*}

 \begin{figure}
\includegraphics[width=0.45 \textwidth]{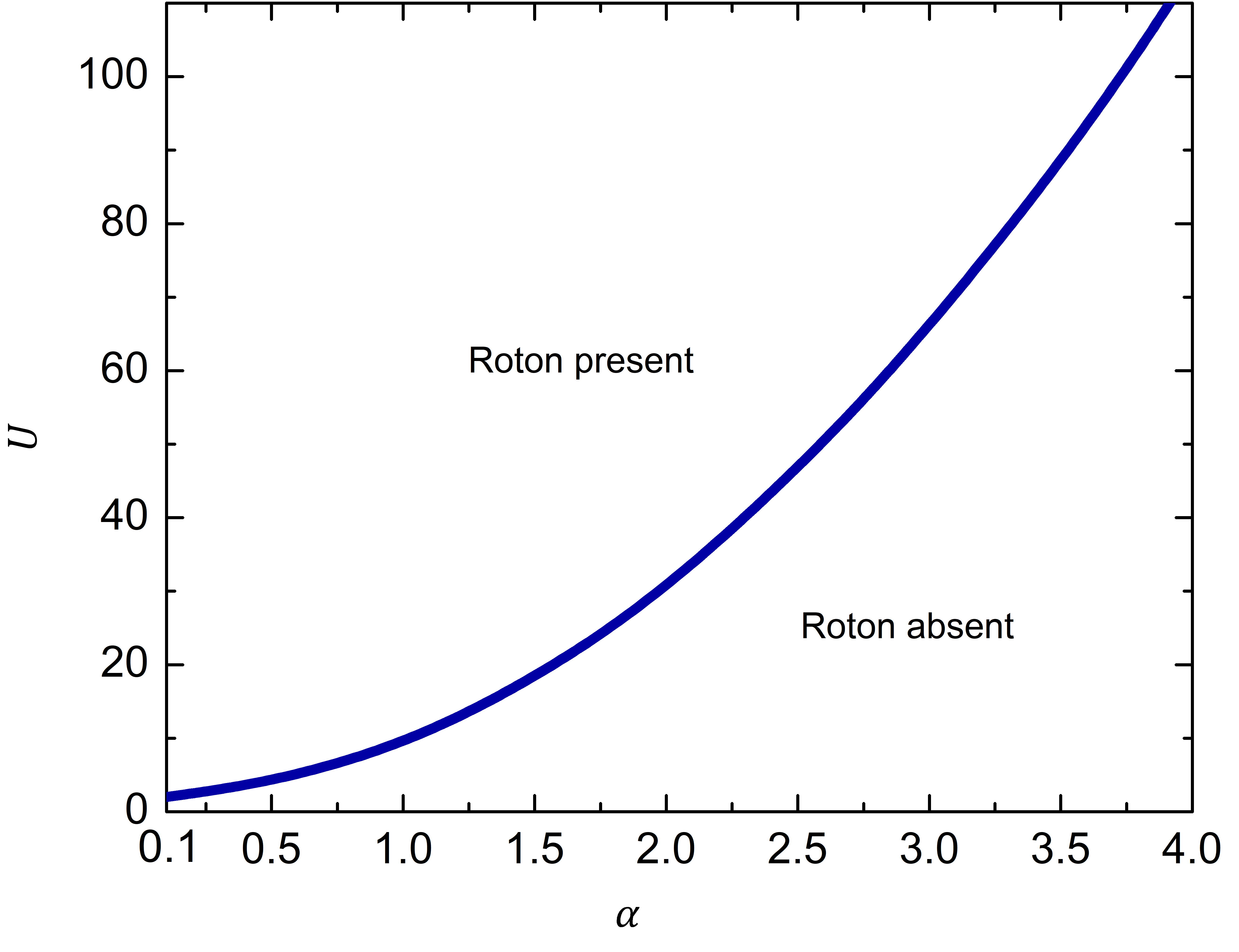}
  \caption{Phase diagram for the appearance of roton excitations in the $(U, \alpha)$ parameter space.  Here $\alpha$ is expressed in units of $\xi^{-1}$ and $U$ is expressed in units of $\rho_c g \, \xi$. The solid curve marks the boundary of roton onset: above the curve the excitation spectrum exhibits a roton minimum (\textit{Roton present}), while below the curve the spectrum remains phonon-like without rotons (\textit{Roton absent}). The boundary was obtained from the condition that the slope (figure \ref {du}) of the excitation spectrum at finite momentum touches zero with no negative slope present, which signals the onset of roton formation.}
  \label{phase}
 \end{figure}

  \begin{figure*}
  \includegraphics[width=5.7cm,height=5.2cm]{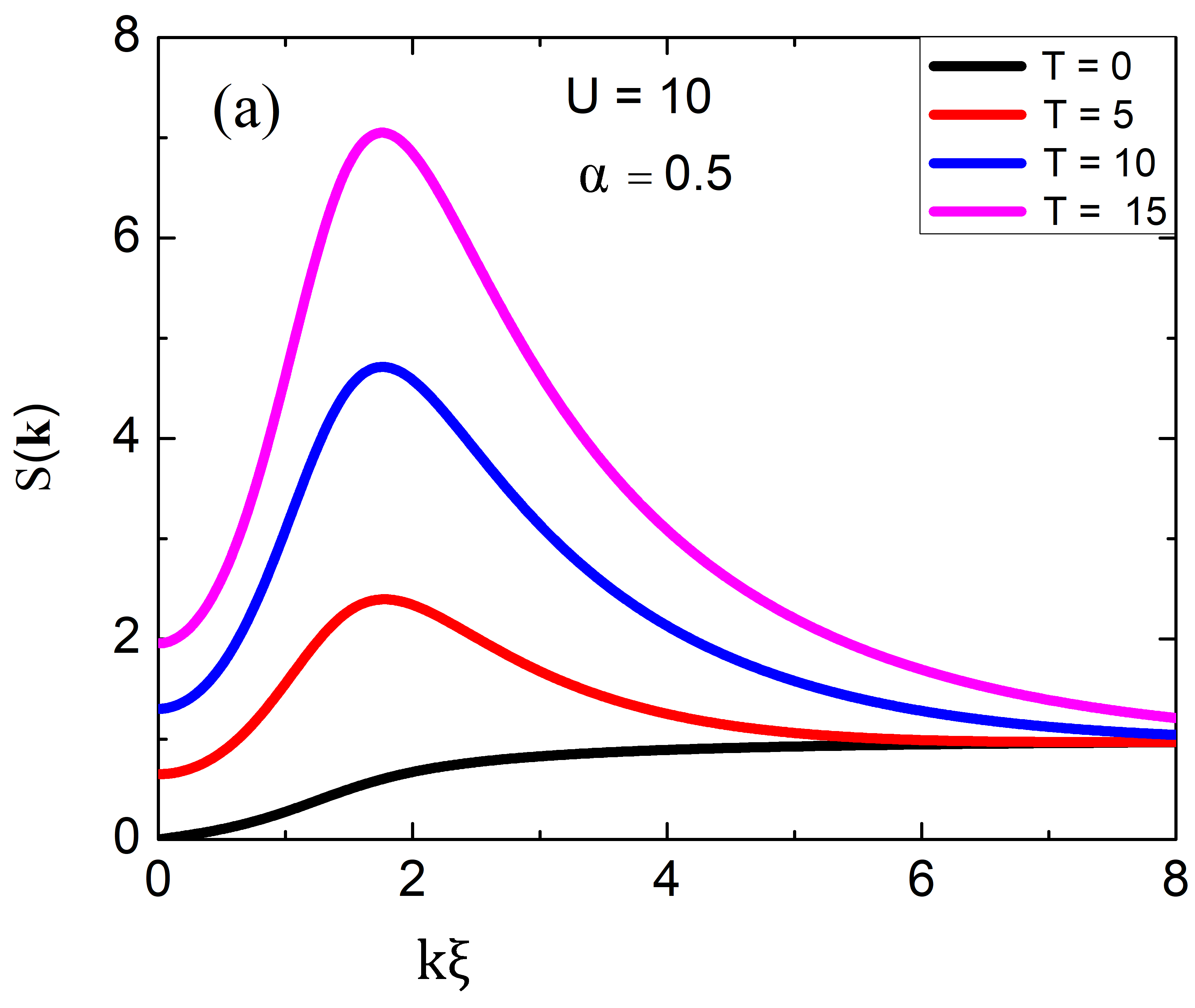}
  \includegraphics[width=5.7cm,height=5cm]  {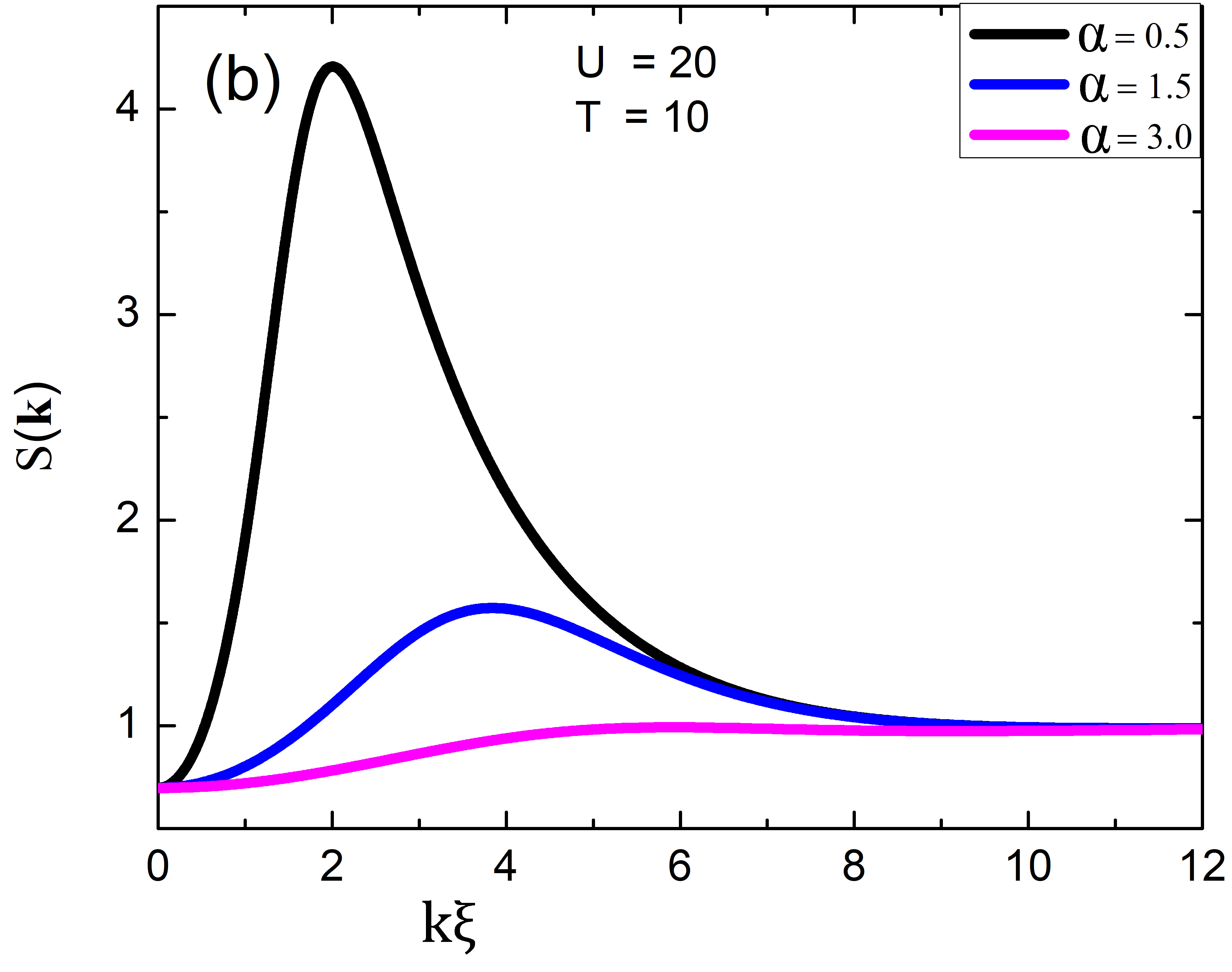}
    \includegraphics[width=5.7cm,height=5cm]  {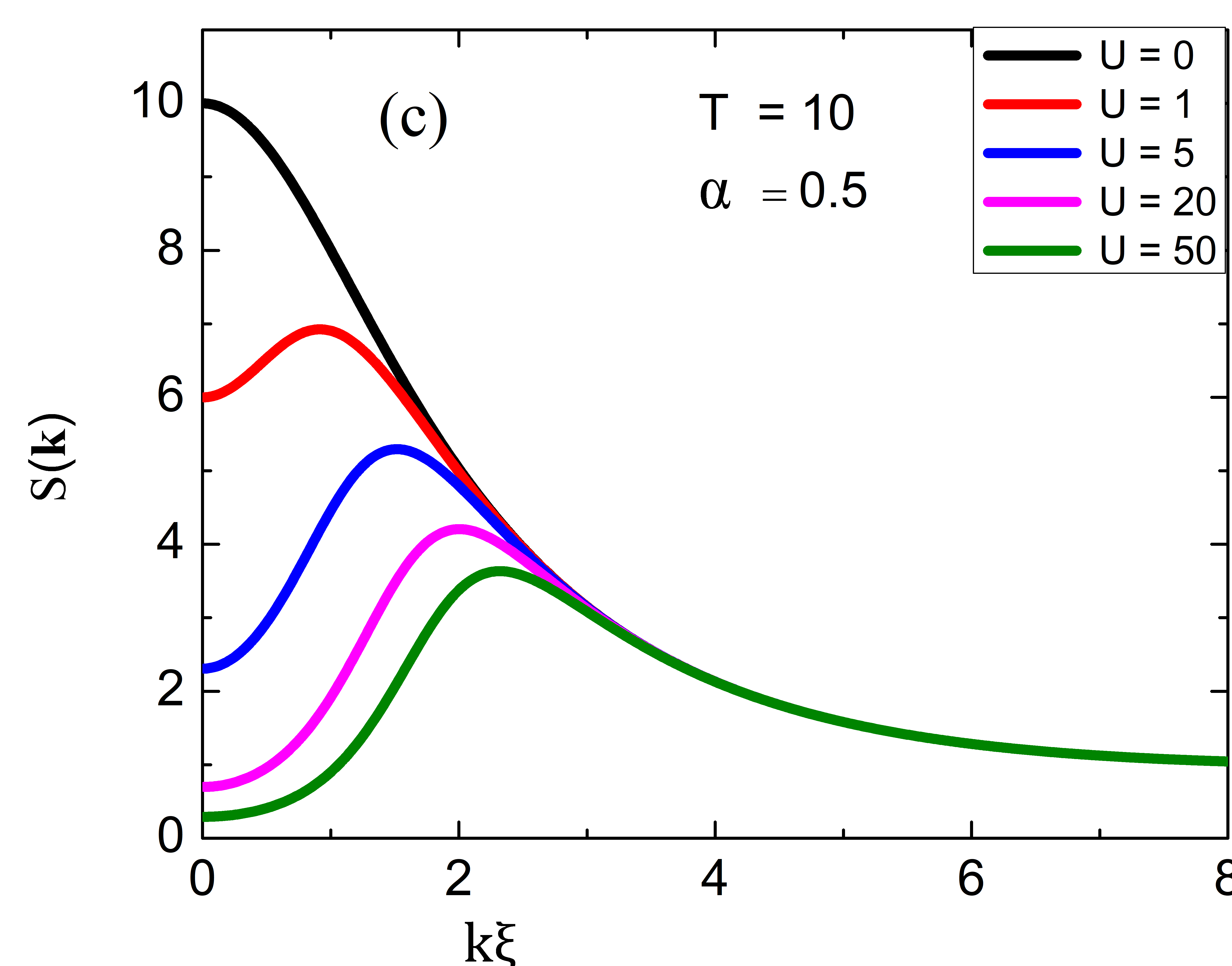}
  \caption{ Structure factor obtained from equation (\ref{sk}). \textbf{[a]} Static structure factor for different temperature (we express temperature in  units of
$ \rho_c g/k_B $) and $U = 10$ and $\alpha= 0.5$. It is studied that the structure factor has a strong temperature dependence \cite{ch_bose2, temp_d} in the lower momentum region. Higher momentum structure factor values tend to unity asymptotically, suggesting a gradual correlation end. \textbf{[b]}  Static structure factor for different $\alpha$ and $U = 20$ and $T = 10$. Peak suggests that the structure factor depends upon the range of PT potential, which is controlled by the parameter $\alpha$. Higher $\alpha$ suggests short-range interaction, and that's why the peak starts to disappear for higher values of $\alpha$ in the low momentum region. At higher momentum, the system is uncorrelated as $S(k)$ tends to unity asymptotically. \textbf{[c] }  Static structure factor for different $U$ and $T = 10$ and $\alpha=0.5$. The peak in the lower momentum region becomes more prominent with increasing $U$. For higher values of momentum, the system is uncorrelated.  $\alpha$ is expressed in units of $\xi^{-1}$ and $U$ is expressed in units of $\rho_c g \, \xi$.} 
   \label{sf}
 \end{figure*}

The field operator of non-condensed atoms can be expressed in terms of plane wave bases as 
\begin{eqnarray*}
 \varphi(\textbf{r})=\frac{1}{\sqrt{\tau}}\sum_{k\ne 0} a_ke^{i\textbf{k}.\textbf{r}} 
\end{eqnarray*}

 where $\tau$ is the volume of the system and $a_k$ is the annihilation particle operator.  The Hamiltonian  can be obtained from equation (\ref{H_nc2}) for a uniform dilute system,
\begin{small}
\begin{eqnarray}
  H = \sum_k \epsilon_k a^\dagger _k a_k + \frac{1}{2} \sum_{k} \mathcal{V}_{T}(k) [2 a^\dagger_ka_k+a_k^\dagger a_{-k}^\dagger
  +a_ka_{-k}],
\end{eqnarray}
\end{small}
where  $\epsilon_k=\frac{\hbar^2k^2}{2m}$ is the free particle energy and
\begin{equation*}
 \mathcal{V}_{T}(k)=\rho_cg(1+ \mathcal{V}_{PT}(k)/g) 
  \end{equation*}
 \begin{equation}
\mathcal{V}_{PT}(k) ={\displaystyle \int } d^3\textbf{r}\mathcal{V}_{PT}(\textbf{r})e^{i \textbf{k} \cdot  \textbf{r} }.
 \end{equation}

where $\mathcal{V}_{PT}(\textbf{r}) = U \cdot 2\alpha/\cosh^2(\alpha\textbf{r})$.   The Fourier transform of the PT potential is evaluated numerically following the definition in ref. \cite{adhikary}. The resulting $\mathcal{V}_{PT}(k)$ tends to zero for large $k$. In the small-$k$ limit, $\mathcal{V}_{PT}(k)$ approaches a finite positive value. We have expressed the Hamiltonian in terms of binary operators, but it is not diagonalized yet. To diagonalize the Hamiltonian we introduce new basis $\{b_k, b_k^\dagger\}$ forms from the existing using Bogoliubov transformation as 
\begin{eqnarray}
  a_k         = u_k b_k - v_k b^\dagger_{-k} \nonumber \\ 
  a^\dagger_k = u_k b^\dagger_k - v_k b_{-k}
\end{eqnarray}
where $u_k$ and $v_k$ are the transformation coefficients. So, the Hamiltonian, in the new basis, becomes

\begin{eqnarray}
 && H = \sum b_k^\dagger b_k \left [ (u_k^2+v_k^2)(\epsilon_k + \mathcal{V}_{T}(k)) - 2 u_k v_k \mathcal{V}_{T}(k) \right ] \nonumber \\
          && + \sum  b_k^\dagger  b_{-k}^\dagger \left [ -u_kv_k (\epsilon_k + \mathcal{V}_{T}(k)) + \frac{1}{2}  (u_k^2+v_k^2) \mathcal{V}_{T}(k) \right ] \nonumber \\
         && + \sum  b_k  b_{-k} \left [ -u_kv_k (\epsilon_k + \mathcal{V}_{T}(k)) + \frac{1}{2}  (u_k^2+v_k^2) \mathcal{V}_{T}(k) \right ] \nonumber \\
          && + \sum \left [ v_k^2 (\epsilon_k + \mathcal{V}_{T}(k)) - u_kv_k\mathcal{V}_{T}(k) \right ].
\end{eqnarray}

The Hamiltonian should be diagonalized in the new basis, so the coefficients of $ b_k  b_{-k}$ and $b_k^\dagger  b_{-k}^\dagger$ must be zero. In addition we have used the normalization condition $u_k^2 - v_k^2 = 1$ to get the transformation coefficients
\begin{eqnarray}
  u_k^2 = \frac{1}{2}\Biggr(\frac{\epsilon_k+\mathcal{V}_{T}(k)}{E_k}+ 1\Biggr) 
\end{eqnarray}
and
\begin{equation}
 v_k^2= \frac{1}{2}\Biggr(\frac{\epsilon_k+\mathcal{V}_{T}(k)}{E_k}-1 \Biggr)
\end{equation} 
\\

where  $E_k= \sqrt{\epsilon_k^2+2\epsilon_k \mathcal{V}_{T}(k)}$.
Putting the expressions of $u_k$ and $v_k$, we get the Hamiltonian as

\begin{equation}
H=\sum_k E_k b_k^\dagger b_k -\frac{1}{2}\sum (\epsilon_k + \mathcal{V}_{T}(k) - E_k).
\end{equation}
The excitation energy is given by,

\begin{equation}
 \frac{ E_k}{\rho_c g} =\frac{1}{2}\xi k[(\xi k)^2+4(1+\mathcal{V}_{PT}(\xi k)/g)]^{1/2}.
\end{equation}

It is convenient to express the excitation energy and momentum as a dimensionless quantity $\Bigr(\frac{E_k}{\rho_cg}$ and $\xi k\Bigr)$ where the healing length of the system is given by $\xi=\frac{\hbar}{\sqrt{m\rho_cg}}$.\\

\textbf{Structure factor:} 

The structure factor is a significant topic to study as it relates to density and quantum fluctuations. It also provides information on the coherence of the system. The structure factor is also helpful in estimating the compressibility of the system. At finite temperature, the structure factor can be expressed as \cite{ch_bose, ch_bose2}

\begin{equation}
S(k) = \frac{\epsilon_k}{E_{k}} \coth (E_k/ 2T).
\label{sk}
\end{equation}

The isothermal compressibility of the condensate can be estimated from the relation
\begin{eqnarray}
  S(k=0) = \rho_c  T \kappa_T
\end{eqnarray}
where $\kappa_T$ is the isothermal compressibility. In the numerical results, we express temperature in units of $\rho_c g/k_B$ where $k_B$ is the Boltzmann constant.

\section*{ Result and discussions}  

We have studied the collective excitations of a uniform Bose gas in the condensed phase using the Bogoliubov approach. Our primary objective is to explore how a finite interaction range between atoms influences the roton mode. In Fig. 1, we present the energy spectra for different values of  $\alpha$, representing the interaction range. Our analysis reveals that the roton mode becomes more pronounced with long-range interactions. As the interaction range decreases (i.e., as $\alpha$ increases), the roton minima become shallower and shift to higher momentum values. Fig. 2 illustrates the energy spectra for varying interaction strengths $U$, with a fixed interaction range of $\alpha = 1$. Without the Pöschl–Teller (PT) interaction, the excitation spectrum exhibits a phonon mode but lacks any roton feature (black line). It is evident that, with increasing interaction strength U, the minima in the spectrum become more pronounced and shift toward higher momentum values. However, the low-momentum excitations continue to display phonon-like behavior.

Fig. 3 illustrates the slope of the excitation spectra. The sound velocity, the slope of the spectra at $k\rightarrow 0$, increases with the interaction strength $U$; however, it remains unaffected by the range of interaction (part [b] of Fig. 3). A negative slope in the excitation spectrum is indicative of the emergence of a roton mode. This roton mode becomes more prominent as the interaction's strength and range increase. In contrast, excitations at large wave vectors are insensitive to the Pöschl–Teller (PT) interaction and are predominantly governed by the contact interaction.

Figs. 4(a) and 4(b) depict the dependence of the roton minima energy $\Delta$ (it is defined as the lowest value of the excitation energy at the roton momentum region) as a function of the parameters $U$ and $\alpha$, respectively. The roton mode disappears for $\alpha> 2.6$ when $U = 50$, as the interaction becomes short-range for larger
values of $\alpha$. Fig. 4(c) presents the sound velocity (it is defined as the slope of the excitation spectrum in the long-wavelength limit, $v_s = \lim_{k  \to 0} \frac{dE_k}{d(k \xi)}$, which corresponds to the phonon velocity obtained from Bogoliubov theory)  variation with $U$. It shows that the sound velocity increases with $U$, whereas it is independent of the range of interaction, as observed from Fig. 1. To characterize further the onset of roton formation (Fig. \ref{phase}), we mapped out a phase diagram 
in the $(U, \alpha)$ parameter space. The boundary was determined from the excitation spectrum by identifying where the slope $dE_k/d(k\xi)$ becomes zero at finite 
momentum with no negative slope present. Above this curve, the system supports a roton minima, whereas below it, the excitation spectrum remains phonon-like without rotons. In Fig. 6, we have presented the structure factor for various interaction parameters and temperatures. Compressibility does not depend on the range of interaction (Fig. 6(b)), but it decreases with the strength of interactions and becomes incompressible at high interaction strength (Fig. 6(c)).

\subsection*{Conclusion}

In this study, we investigated the collective excitations of a uniform Bose gas with 
finite-range P\"oschl–Teller (PT) interaction in addition to the usual contact interaction, 
using the Bogoliubov approach. Our analysis demonstrates that:

\begin{itemize}
    \item The roton minima strongly depend on both the range and the strength of the PT interaction. It becomes more pronounced for long-range interactions, but it does not appear when the interaction range is very small.
    
    \item The sound velocity increases  with the interaction strength $U$,  but remains essentially independent of the interaction range parameter $\alpha$.
    
    \item The compressibility of the condensate decreases as the interaction strength increases, and is insensitive to the interaction range.
\end{itemize}

These results highlight how the tunable parameters of the PT potential provide a versatile framework 
to study the emergence and suppression of roton excitations in Bose–Einstein condensates.

\end{document}